\documentclass[12pt,epsfig,letterpaper]{article}
\ifx\pdfoutput\undefined
\usepackage[dvips,bookmarks]{hyperref}	
\else
\usepackage{hyperref}	
\fi
\def\hhref#1{\href{http://arxiv.org/abs/hep-th/#1}{hep-th/#1}} 
\def\mhref#1{\href{mailto:#1}{#1}}		

\catcode`@=11
\def\secteqno{\@addtoreset{equation}{section}%
\def\theequation{\thesection.\arabic{equation}}}
\catcode`@=12
\secteqno
\newcommand{\be}{\begin{equation}}
\newcommand{\ee}{\end{equation}}
\newcommand{\bea}{\begin{eqnarray}}
\newcommand{\eea}{\end{eqnarray}}

\def\half{{\textstyle{1\over{\raise.1ex\hbox{$\scriptstyle{2}$}}}}}
\catcode163=13 \def\itm{\relax\ifmmode\to\else\itemize\fi}

\newcommand{\mathbb}{\mathbf}   

\begin{document}
\thispagestyle{empty}

\hfill CINVESTAV-FIS-15/04

\hfill April, 2004

\vskip 20mm

 \begin{center}
 {\Large{\bf A Reduced Model of Noncommutative \\ S-brane Spectrum}}
 \vskip 6mm
 \medskip
 \vskip 10mm
 {\large H. Garc\'{\i}a-Comp\'ean$^{\ast}$~and~ J. Gonz\'alez-Beltr\'an$^\dagger$}

\parskip .15in
{ 
 Departamento de F\'{\i}sica} \\
 Centro de Investigaci\'on y de Estudios Avanzados del IPN\\
 Apdo. Postal 14-740, 07000, M\'exico D.F., M\'exico \\
 $^\ast${\small{ e-mail address:\ \mhref{\tt compean@fis.cinvestav.mx} }}\\
 $^\dagger${\small{ e-mail address:\ \mhref{\tt jbeltran@fis.cinvestav.mx} }}\\

 \medskip
 \end{center}
\vskip 10mm
\begin{abstract}
CFT construction of S-branes describing the rolling and bouncing tachyons is analyzed in the
context of a $\theta$-noncommutative deformation of minisuperspace. Half s-brane and s-brane in
the noncommutative minisuperspace are analyzed and exact analytic solutions, involving the
noncommutative parameter $\theta$ and compatible with the boundary conditions at infinity, are
found.  Comparison with the usual commutative minisuperspace is finally performed.  
\end{abstract}

\vskip 3cm
\setcounter{page}{1}
\parskip=7pt
\newpage

 \section{Introduction}

\setcounter{figure}{0}

 It is common knowledge that the presence of tachyonic modes in 
 a quantum field theory signals the existence of an unstable vacuum.
 In the past, ``annoying" tachyon states would be responsible for the
 discredit of their own embracing field theory which was then doomed
 to oblivion unless some cure could be implemented. In the case of
 string theory for instance, the mechanism credited for getting rid of
 this unappealing circumstance was the remarkable discovery of 
 supersymmetry. Not surprisingly, it has been precisely string theory
 which has taught us over the past few years that there is also much
 to be learned in the study of the long time eluded tachyon. As a result
 we have witness an increasing interest on this reach and extensive field
 of research (for reviews on different subjects concerning tachyons,see
\cite{revs}).
 
 A particular case of a system with tachyonic excitations in string theory
 which later on grew in importance due to its relevance in time-dependent
 string backgrounds (see below) is that of a coincident D-brane  
 anti D-brane pair. It was realized in \cite{dAd_Tach} that in the NS sector
 ground state of this system survives the GSO projection giving rise to 
 a complex scalar field with negative squared mass for Type II
 theory, this renders the system unstable.  As shown by Sen \cite{sen} 
 a tachyonic kink solution on this brane anti-brane pair can be identified 
 with a stable non-BPS D-brane of the same theory in one lower dimension. 
 On the other hand, he also showed how BPS D-branes can be associated with
 tachyonic kink solutions on non-BPS D-branes on one higher dimension, 
 effectively uncovering a web of descent relations between BPS and non-BPS
 D-branes of Type II string theories, opening the way to the description
 of D-brane charges in terms of elements of K-theory \cite{k-th}.
 
 The above kink solutions that interpolate between two minima of the
 tachyon potential acquired relevance in the context of time-dependent
 string backgrounds when Strominger and Gutperle \cite{Gutperle:2002ai}
 adapted Sen's argument and showed that when the interpolation occurs
 in the timelike direction we are effectively describing objects which
 are localized on a spacelike hypersurface. These are known as S-branes
 or Spacelike branes and can be thought of as the description of the 
 time-dependent processes involving the creation and subsequent decay of
 unstable D-branes in string theory. Also, because of their own nature, 
 S-brane solutions in string and supergravity theories have attracted
 much attention towards the possible cosmological interpretations. For
 recent works on S-branes see \cite{Ivashchuk:2004zb} throughout 
 \cite{Gutperle:2003xf}, 
 research in the cosmological aspects include \cite{Leblond:2004dm}
 throughout \cite{Ohta:2003pu}.

 Noteworthy, in \cite{Sen:2002nu} Sen explores the dynamics of the tachyon from
 a string field theory point of view. He showed that the worldsheet action
 of the boundary conformal field theories associated with the classical 
 time-dependent solutions describing the motion of unstable bosonic D-branes can
 be taken to be
\be\label{worldsheet} 
S = -\frac{1}{4\pi} \int_{\Sigma_2} d^2 \sigma \partial^a X^\mu \partial_a 
 X_\mu  +  \lambda  \int_{\partial \Sigma_2} d\tau \cosh X^0(\tau), 
\ee
 (we use $\alpha^\prime = 1$)
 and he gave an analogous treatment for the superstring in \cite{Sen:2002in}.
 In the formal limit of weak coupling constant $g_s \rightarrow 0$ the action
 (\ref{worldsheet}) actually corresponds to a CFT construction of the spacelike
 brane. A related boundary interaction of 
\be\label{half} 
  \frac{\lambda}{2}  \int_{\partial \Sigma_2} d\tau \exp(X^0), 
\ee
 was discussed in \cite{Strominger:2002pc} motivated in part by the
 resemblance of this system with the boundary Liouville theory which has been
 extensively studied before \cite{BLT} (eventually this led to the 
 construction of timelike boundary Liouville theory \cite{Gutperle:2003xf}). 
 Following the usual terminology we will refer to (\ref{worldsheet}) as the
 {\it bouncing} tachyon or s-brane and to (\ref{half}) as the {\it rolling} tachyon
 or half s-brane.

 Our main interest in this work will be focused on the minisuperspace
 approximation to these systems first introduced in this context by
 Strominger in \cite{Strominger:2002pc} in analogy to the ordinary Liouville
 theory \cite{ordmini}. In these reduced models the approximation consists
 in considering only the zero-mode dependence of the boundary interactions.
 Recently a detailed analysis of the minisuperspace energy spectrum of these
 models was performed in \cite{Fredenhagen:2003ut} (for further work in the context of the 
minisuperspace approach, see \cite{hikida}) and the main task in 
 this note will be to extend their results to a {\emph {deformed}} minisuperspace
 on which we shall introduce a noncommutative parameter between the minisuperspace
 variables in the spirit of \cite{Seiberg:1999vs}{\footnote {
 For an approach to the full space noncommutative
 tachyon see for instance \cite{Gopakumar:2000zd}.}}.
 The main motivation comes 
 from an analogous treatment in the context of quantum gravity in reference
 \cite{Garcia-Compean:2001wy} wherein it is introduced a noncommutativity
 between the variables of the reduced quantum cosmology model of the 
 Kantowski-Sachs metric. In section \ref{qgrav} we will give a brief
 review of this ideas introducing a novel example of an exact solution
 to the noncommutative Wheeler-DeWitt equation for the standard massless
 scalar field minimally coupled. It is worth to mention that a somewhat different approach
involving the minisuperspace formulation of quantum S-branes has been worked out and will appear
elsewhere \cite{gerardo}.

 Nevertheless, we have to point out that the noncommutativity considered here
 in the context of tachyon dynamics is of a different nature than that of
 quantum cosmology minisuperspace models, since usually the minisuperspace
 variable $x^0$ 
 which parametrizes the dependence of the boundary interaction terms in the
 reduced forms of the 
 actions above is indeed the timelike parameter, and we will assume a non-zero
 noncommutative parameter between the commutators $[x^\mu,x^\nu]$, including
 those of $x^0$ with its spacelike counterparts. Actually in the literature it
 is often assumed that the noncommutativity is present only in the spacelike
 directions of the system under consideration, the reasons for this are fairly
 obvious in that common issues of causality have to be considered, in fact it
 is known that fundamental changes have to be implemented, for instance, in the  
 formulation of ordinarily quantum mechanics if a noncommutative timelike 
 parameter is included \cite{Li:2001vc}.  In spite of these facts, we
 will push on and continue with our assumptions since we believe that ideas of
 this kind deserve a closer examination, as have been shown in the past in other 
 contexts{\footnote {For instance, the study of timelike dualities has proved
 worth of attention in the context of the search for a formulation of a 
 dS/CFT duality \cite{Hull:1998vg}.}}. 

 Keeping these remarks in mind throughout the rest of this paper, we will
 introduce a deformation of the minisuperspace models above in sections 
 \ref{rollingsecc} and \ref{sbranesecc} for the half s-brane and the s-brane 
 respectively. We will see how the resulting effective equation for the
 corresponding energy spectrums have analytic solution without the need for
 an expansion in the noncommutativity parameter $\theta$. The resulting 
 eigenstates of the corresponding models acquire a shift in the timelike 
 parameter proportional to $\theta$. For the purpose of visualizing the 
 effect of noncommutativity, wave packets with suitable weighing functions
 are constructed and analyzed for special cases. In general the $\theta$-effect
 shows up as an enhancement of the localization in the timelike direction
 of the wave packet, apart from an overall shift. 
 Finally in section \ref{sum} we summarize and comment on our results.


 \section{Overview of the Noncommutative Wheeler-DeWitt \\ Equation}\label{qgrav}

 Quantum Gravity is an old subject of theoretical physics. It grew in the late 
 1930's from the desire to emulate the successes which the then new quantum field
 theory had had with regards the beginning of quantum electrodynamics. The feeling at that time
was that only technical difficulties 
 remained to be surpassed to achieve the desired goal. At present, a full
 description of a quantum theory of the gravitational field remains a (yet to 
 be found) eagerly searched milestone of contemporary knowledge. 
 The simplest model of quantum gravity describes the quantum behavior of
 3-space regarding it as a time-varying geometrical object. The quantum states
 are usually chosen in the so called metric representation in which the wave
 function of the (entire) universe $\Psi$ becomes a functional of the metric
 components $g_{\mu\nu}$ and the momenta become functional differential 
 operators with respect to the metric components. Consistency with the canonical
 constraints then leads to the well known Wheeler-DeWitt equation for the 
 functional $\Psi$ which then depends solely on the spatial components $g_{ij}$ 
 of the metric, and the space of all consistent 3-geometries is denominated
 superspace. Of course most of the analysis of the Wheeler-DeWitt equation 
 deal with a reduced form of this superspace known as minisuperspace, in which
 almost all of the degrees of freedom of the gravitational field are `frozen out'. 
 As our example of this minisuperspace techniques we shall consider metrics 
 which in appropriate units read
\be\label{metric}
ds^2 = N^2(t) dt^2 + a^2(t) d\Omega^2_3,
\ee
 where $N(t)$ is known as the lapse function, $a(t)$ is the scale factor 
 and $d\Omega^2_3$ is the metric of a three-sphere of unit radius. Actually 
 the solutions of the Wheeler-DeWitt equation are independent of $N$ and $t$
 since the wave function $\Psi(a,\phi)$ is a function solely of $a$ and possible
 matter fields denoted by $\phi$. We shall consider only a massless scalar 
 field $\phi$ minimally coupled to the gravitational filed. In this case the
 Wheeler-DeWitt equation with the standard factor ordering prescription reads
 \cite{Hawking:in}:
\be\label{WdW}
\left[ a \frac{\partial}{\partial a} a \frac{\partial}{\partial a}  - 
\frac{\partial^2}{\partial \phi^2} - a^4 \right] \Psi(a,\phi) = 0.
\ee
 Solutions to this equation are readily found and their interpretation range
 from wave functions corresponding to classical Lorentzian-Friedmann 
 universes to wormholes for appropriate linear combinations satisfying 
 certain boundary conditions \cite{Hawking:in}. 

 The proposal of  \cite{Garcia-Compean:2001wy}    was to introduce a 
 noncommutativity between the minisuperspace
 variables, in this case $a$ and $\phi$, so as to mimic the spacetime 
 noncommutativity motivated by string theory. With this in mind, 
 we will assume a non-zero commutator between $a$ and $\phi$ is given by
\be\label{qconmmu}
[a,\phi] = i \theta,
\ee
 where $\theta$ is the noncommutativity parameter.

 The standard procedure now is to implement this noncommutativity in the
 the theory through a Moyal deformation of the usual product of functions.
 Thus the noncommutative Wheeler-DeWitt equation which corresponds to 
 (\ref{WdW}) reads
\be\label{noncommWdW}
\left[ a \frac{\partial}{\partial a} a \frac{\partial}{\partial a}  - 
\frac{\partial^2}{\partial \phi^2} - a^4 \right] * \Psi(a,\phi) = 0,
\ee
 where all products of functions are with respect to the Moyal star
 product $*$ defined by
\be\label{moyalstar}
f(a,\phi) * g(a,\phi) := f(a,\phi) e^{i(\theta/2)\left( \overleftarrow{\partial_a} 
\overrightarrow{\partial_\phi} -  \overleftarrow{\partial_\phi} 
\overrightarrow{\partial_a}   \right)}  g(a,\phi). 
\ee
 Thus for instance $(a \frac{\partial}{\partial a} a \frac{\partial}{\partial a})
 * \Psi(a,\phi)$ stands for $ a*\frac{\partial}{\partial a} \left( 
 a*\frac{\partial \Psi(a,\phi)}{\partial a} \right)$. Now we make use of 
 the identity \cite{Bigatti:1999iz}:
\be\label{star-pot}
V(a,\phi)*\Psi(a,\phi) = V(a + \frac{1}{2} i \theta \frac{\partial}{\partial \phi},
 \phi -  \frac{1}{2} i \theta \frac{\partial}{\partial a})  \Psi(a,\phi), 
\ee
 where the product of functions in the rhs is the usual product of functions. Upon 
 substitution of this expression on (\ref{noncommWdW}) the advantage of
 the chosen model becomes apparent since no $\phi$-terms arise in the resultant
 equation, only derivatives with respect to $\phi$. This makes the resultant
 equation separable. Choosing the $\phi$-dependence of the harmonic solution
 $\Psi(a,\phi) = e^{i k \phi} \psi(a),$ the resulting ordinary differential
 equation for the $a$-dependence is:
\be\label{a-eq}
\psi^{\prime\prime}(a) + \frac{2}{2a-k\theta} \psi^\prime(a) 
+ \frac{16 k^2 - (k\theta-2a)^4}{4(2a-k\theta)^2} \psi(a)  = 0, 
\ee
 where prime's denote differentiation with respect to $a$. The solutions to 
 this equation are modified Bessel functions of the first kind $I_n(z)$ of
 imaginary order $n$, however we prefer to analyze linear combinations of 
 this solutions which are well behaved (finite) for large $a$, thus our 
 complete solutions are of the form:
\be\label{qsol}
\Psi^\pm_k(a,\phi) = e^{i k \phi} K_{\pm \frac{ik}{2}} \left(
  \frac{1}{8} (k\theta - 2a)^2 \right),
\ee
 where $K_n(z)$ are the modified Bessel functions of the second kind. Following
 \cite{Garcia-Compean:2001wy}, we construct the special gaussian weighted wave
 packet
\be\label{qgauss}
\Omega(a,\phi) = {\cal {N}} \int_{-\infty}^\infty 
                 e^{-\tau(k - \eta)^2} \Psi^+_k(a,\phi) dk,
\ee
 so as to be able to visualize the effect of the noncommutativity parameter
 $\theta$ in the shape of the probability distribution for this packet. Here
 ${\cal {N}}$ is a normalization constant and $\tau$ and $\eta$ are adjustable
 parameters. The integral is performed numerically and in figure 1
 we show a comparison between the case of zero noncommutative parameter ($\theta =0$) and the 
 non-zero one. There we show a slice of the probability distribution $|\Omega|^2$
 as a function of $a$ for the value $\phi=0$. As can be seen from the figure, the 
 most notable effect of noncommutativity is to induce the formation of a local
 maximum apart from the absolute maximum encounter in the $\theta=0$ case. This is
 an example of the phenomenon discovered in \cite{Garcia-Compean:2001wy}    
 which was interpreted there as the possibility of tunneling among different
 states of the universe. 

 This completes our review of noncommutative quantum gravity and we now 
 proceed to implement some of these ideas to the context of the bouncing and rolling tachyons.


\begin{figure}\label{fig:qgravfigone}
\let\picnaturalsize=N
\def\picsize{15.0 cm}
\def\picfilename{qgravfig.eps}
\ifx\nopictures Y\else{\ifx\epsfloaded Y\else\input epsf \fi
\let\epsfloaded=Y
\centerline{\ifx\picnaturalsize N\epsfxsize \picsize\fi \epsfbox{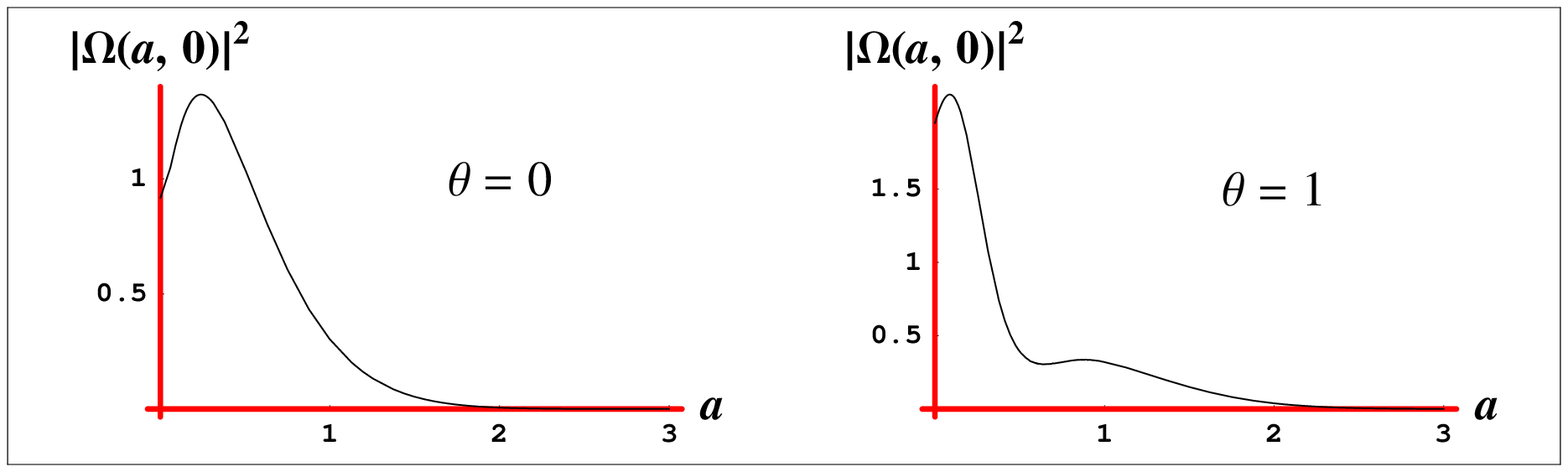}}}\fi
{\small {Figure 1  Probability distributions for the wave packet $\Omega(a,\phi)$ as a function
     of $a$ for the special value $\phi = 0$. The left graph corresponds to zero noncommutative
     parameter and the right one to $\theta=1$. The values of the remaining parameters are 
      $\tau=\eta=1$.}}
\end{figure}

 \section{The Half S-brane}\label{rollingsecc}

 For the half s-brane the worldsheet action is (\ref{worldsheet}) but 
 boundary interaction (\ref{half}) is given by:
\be\label{half-sbrane}
S = -\frac{1}{4\pi} \int_{\Sigma_2} d^2 \sigma \partial^a X^\mu \partial_a 
 X_\mu  +  \frac{\lambda}{2}  \int_{\partial \Sigma_2} d\tau \exp(X^0). 
\ee
 As it was mentioned before, the minisuperspace approximation concerns only the
 zero modes of the fields.
 After the usual mode expansion of the open string the effective action
 is given by
\be\label{eff}
 S = \int d\tau \left\{ -\frac{1}{4} \dot{x}^\mu \dot{x}_\mu  
     + (N-1) +  \lambda \exp x^0 \right\},
\ee
 where it is assumed that $x^\mu = x^\mu(\tau)$, $x^\mu$ being the zero modes 
 of the embedding fields $X^\mu(\sigma,\tau)$, and $N-1$ is the oscillator 
 contribution. To proceed further, it is useful to examine the hamiltonian 
 constraint, this can be 
 cast in the form of a Klein-Gordon type equation for the minisuperspace
 wave function $\psi(x)$ \cite{Strominger:2002pc}:
\be\label{klein-gordon}
\big[\partial^\mu \partial_\mu -  \lambda \exp x^0 - (N-1)\big] \psi(x) = 0. 
\ee 
 At this point we introduce commutation relations for the minisuperspace
 variables as follows
\be\label{conmmutators}
[x^\mu,x^\nu] = i \theta^{\mu \nu},
\ee
 where $\theta^{\mu\nu}$ is antisymmetric and for simplicity we will take it
 to be constant: $\theta^{\mu\nu} = \theta \varepsilon^{\mu \nu}$ for $\mu < \nu$. As is well
known,
 we can implement this noncommutativity through a deformation of the
 minisuperspace product of functions by using the Moyal product:
\be\label{moyal}
f(x) * g(x) = f(x) \exp\left\{  -\frac{i}{2} \overleftarrow{\partial_j} 
        \theta^{jk} \overrightarrow{\partial_k}  \right\}  g(x)
\ee
 (we use Greek letters to denote lorentzian indices raised and lowered with
 metric $\eta^{\mu\nu}$ and Latin letters for euclidean indices, here $j,k
 = 0,1, \ldots, d-1$ as well as $\mu, \nu$ where $d$ is the spacetime 
 dimension and $\theta^{jk}$ is defined as above.)
 In this manner, the noncommutative version of (\ref{klein-gordon}) becomes
\be\label{noncomm-eq}
\big[\partial^\mu \partial_\mu -  \lambda \exp x^0 - (N-1)\big] * \psi(x) = 0. 
\ee
 For our purposes, we rephrase the star product in terms of the well
 known formula \cite{Bigatti:1999iz}
\be\label{formula}
f(x) * g(x) = f\left( x^j - \frac{i}{2} \theta^{jk} \partial_k \right) g(x).
\ee
 By making use of this expression, Eq. (\ref{noncomm-eq}) becomes
\be\label{inter1}
\left(\partial^\mu \partial_\mu - (N-1) -  \lambda \exp\left\{ 
       x^0 - \frac{i}{2} \theta \sum_{k \neq 0} \partial_k  \right\} \right) 
         \psi(x)  = 0.
\ee
 We observe now that with the decomposition in plane waves: 
\be\label{ansatz}
\psi(x) = \exp \big( i {\bf p} \cdot {\bf x}  \big) \phi(x^0),
\ee
 the above equation actually separates, yielding
\be\label{rolling}
\left(  \partial_{x^0}^2 + \lambda e^{\frac{\theta}{2} k} e^{x^0} 
     +  \omega^2  \right) \phi(x^0) = 0,
\ee
 where $\omega^2 = p^2 + N - 1$ and $k = \sum_{j \neq 0} p_j$. 
 In this way we are led to identify the
 {\emph {induced}} hamiltonian due to noncommutativity as the operator
\be\label{rollingHam}
H_{rt} = \partial_{x^0}^2 + \lambda e^{\frac{\theta}{2} k} e^{x^0} 
     +  \omega^2.  
\ee
 We now proceed to analyze $H_{rt}$ in usual terms. This problem was essentially
 solved in reference \cite{Kobayashi:1996kg} (see also  \cite{Fulop:1995di})
 motivated by different reasons than the present one and for clarity of the
 exposition we will review the arguments there in some detail. The first thing we ought to
 state clear is the domain $D(H_{rt})$ of definition of this hamiltonian. The natural
 choice is to require $H_{rt}$ to be an operator defined on the Hilbert space of
 square integrable functions on the whole real line with the standard inner
 product of functions. Thus $x^0$ takes values on the interval $(-\infty,\infty)$
 and the state vectors on this Hilbert space are functions $\psi$ such that
\be\label{squarei}
||\psi||^2 =\langle \psi | \psi \rangle = \int_{-\infty}^\infty \psi^*(x^0) \psi(x^0) dx^0  <
\infty.
\ee
 For future convenience we shall consider the more general class of
hamiltonians
\be\label{Hams}
H = \partial_{x^0}^2 + V(x^0),  
\ee
 where $V(x^0)$ is a real valued smooth function of $x^0$. We will see that for
 the potentials $V$ that we need to deal with the resultant hamiltonians defined
 with such a domain as above are not self-adjoint and we will have to find 
 appropriate self-adjoint extensions of them. At this point it is useful to
 make the change of variable
\be\label{deft}
t = e^{x^0},
\ee
 upon which the hamiltonian reads
\be\label{Hamt}
H = t \frac{d}{dt} t \frac{d}{dt} + V(t).      
\ee
 Now the state vectors are defined on the interval $t \in (0,\infty)$ with scalar product
\be\label{inner}
\langle \psi | \phi \rangle = \int_0^\infty \frac{dt}{t} \psi^*(t) \phi(t),   
\ee
 where the functions have compact support on $(0,\infty)$. Note the induced measure
 on the inner product, because of this, functions belonging to the domain of $H$
 must vanish as $t$ approaches to zero.  Suppose that $\psi,\phi \in D(H)$,
 since $V$ is real-valued, with a straightforward partial integration we can show that
\be\label{issymm}
\langle \phi | H \psi \rangle =  \langle H \phi |  \psi \rangle + 
 \left[ \phi^* t \frac{d\psi}{dt} - \frac{d\phi^*}{dt} t \psi \right]  {\Bigg |_0^\infty}.
\ee
 From this expression we can find a suitable domain in which $H$ becomes a 
 symmetric operator, namely
\be\label{symm}
D(H) := \left\{ \psi | \psi \in L^2(0,\infty), H \psi \in L^2(0,\infty),
 \lim_{t \rightarrow 0^+, \infty}  t \frac{d\psi(t)}{dt} = 0 
\right\}.
\ee
 On the other hand suppose that $\psi$ belongs to the domain of the adjoint
 operator $H^*$, then for all $\phi \in D(H)$ we have
\be\label{adjoint}
\bigg\langle t \frac{d}{dt} t \frac{d}{dt} \phi + V(t)\phi | \psi \bigg\rangle  
  = \langle \phi | H^* \psi \rangle. 
\ee
 If we further assume that the functions in the domain of $H^*$ are well behaved 
 (finite) as $t$ approaches $0$ or $\infty$, then because of the properties of
 the functions that belong to $D(H)$ we must have
\be\label{act}
\bigg\langle \phi | t \frac{d}{dt} t \frac{d}{dt} \psi + V(t)\psi \bigg\rangle  
  = \langle \phi | H^* \psi \rangle. 
\ee
 Thus $H^*$ as a differential operator acting on the functions of its domain
 has the same form as $H$. This is very useful in the computation of the deficiency indices
 $(n_+,n_-)$ of the symmetric operator $H$, which are the dimensions of the kernel 
 ${\cal{K}_\pm}$ of the respective operators $(H^* \pm i)$. First note that $H$ 
 commutes with complex conjugation since $V$ is real, this tells us that its deficiency
 indices are equal $n_+=n_-:=n$, and therefore $H$ is self-adjoint or admits
 self-adjoint extensions depending of whether or not $n$ is equal to zero. 
 
 In order to calculate $n$ we have to particularize the discussion to $H_{rt}$
 defined by Eq.  (\ref{rollingHam}). The solutions of the equations $H^*_{rt} \psi = \pm i
\psi$
 are Bessel functions of the first kind of complex order, two independent solutions
 for each equation. However for each sign there is only one square integrable 
 solution, this can be checked easily from the asymptotics of the Bessel functions.
 Thus the respective kernel subspaces are
\be\label{kerp}
{\cal{K}_+} = \left\{ \beta J_{2\sqrt{i-\omega^2}} \left[ 2\sqrt{\lambda} e^{\frac{k\theta}{4}}
                t^{\frac{1}{2}} \right] | \beta \in {\mathbb{C}} \right\},
\ee
\be\label{kerm}
{\cal{K}_-} = \left\{ \beta J_{2i\sqrt{i+\omega^2}} \left[ 2\sqrt{\lambda} e^{\frac{k\theta}{4}}
                t^{\frac{1}{2}} \right] | \beta \in {\mathbb{C}} \right\},
\ee
 and therefore the deficiency indices of $H_{rt}$ are $(1,1)$, confirming that it is
 not self-adjoint. To construct the self-adjoint extensions we need to find the partial
 isometries $U$ from a set $I(U) \subseteq {\cal{K}_+}$ into ${\cal{K}_-}$. An operator $U$
 is an isometry if for all $\psi$ belonging to the Hilbert space in question we
 have $||U \psi || = || \psi ||$. It is a partial isometry if it is an isometry when
 restricted to states not belonging to its kernel. To construct the possible operators
 $U$ we first give the states that span ${\cal{K}_\pm}$, to this end we make use of the
 relation
\be\label{lommel}
\int_a^b \frac{dz}{z} J_k(z) J_l(z) = \frac{1}{k^2 - l^2} \left[ 
   z \left( J_k(z) \frac{d}{dz} J_l(z) - J_l(z) \frac{d}{dz} J_k(z)  
\right) \right]{\Bigg|_a^b}. 
\ee
 For $a \rightarrow 0$ and $b \rightarrow \infty$ we can profit from the asymptotics
 expressions for the Bessel functions
\be\label{asymbi}
J_k(z) {\longrightarrow_{z \rightarrow 0}} \qquad \sqrt{\frac{2}{\pi z}} 
       \cos[z - \frac{\pi}{2}(k+\frac{1}{2})],
\ee
\be\label{asymb0}
J_k(z) {\longrightarrow_{z \rightarrow 0}}  \qquad \frac{1}{\Gamma(k+1)} (\frac{z}{2})^k , 
\ee
\be\label{asymbYi}
Y_k(z) {\longrightarrow_{z \rightarrow \infty}} \qquad \sqrt{\frac{2}{\pi z}} 
       \sin[z - \frac{\pi}{2}(k+\frac{1}{2})],
\ee
\be\label{asymbY}
Y_k(z) {\longrightarrow_{z \rightarrow 0}}  \qquad - \frac{1}{\pi} \Gamma(k) 
  (\frac{z}{2})^{-k}, \ \ \ \ \ ({\textrm{Re}} k>0) 
\ee
 where $k\in {\mathbb{C}}$, ${\textrm{arg}} |z|<\pi$ and where we have written the expressions
 for the asymptotic behavior of the Bessel function of the second kind $Y_k(z)$
 for future use. Using this expressions we find
\be\label{lommelnorma}
\int_0^\infty \frac{dz}{z} J_{k^*}(z) J_l(z) = - \frac{2}{\pi} \frac{1}{(k^* + l)(k^*-l)}
   \sin[\frac{\pi}{2} (k^*-l)], 
\ee
 where we have used $J^*_k(z) = J_{k^*}(z)$ for $z\in {\mathbb{R}^+}$.  In particular for
 $k=l$ we have
\be\label{norma}
||J_k(z)||^2  =  \frac{1}{2 {\textrm{Re}}(k)} 
                 \frac{\sinh(\pi {\textrm{Im}}(k))}{\pi {\textrm{Im}}(k)}.
\ee
 With this preliminaries we can write the normalized states $\phi_\pm$ which span
 ${\cal{K}_\pm}$ respectively
\be\label{normvect+}
\phi_+(t) = {\cal {N}} J_{2\sqrt{i-\omega^2}} \left[ 2\sqrt{\lambda} 
         e^{\frac{k\theta}{4}} t^{\frac{1}{2}} \right]
\ee
and
\be\label{normvect-}
\phi_-(t) = {\cal {N}} 
         J_{2i\sqrt{i+\omega^2}} \left[ 2\sqrt{\lambda} e^{\frac{k\theta}{4}}
                t^{\frac{1}{2}} \right], 
\ee
 where ${\cal {N}}$ is the normalization factor of both states since
 it turns out to be the same due to the fact that $2i\sqrt{i+\omega^2} = 
 (2\sqrt{i-\omega^2})^*$, namely
\be\label{normfact}
{\cal {N}} :=  
     \left( {\textrm{Re}} (2\sqrt{i-\omega^2}) \frac{{\pi {\textrm{Im}} 
(2\sqrt{i-\omega^2})}} 
      { {\sinh(\pi {\textrm{Im}} (2\sqrt{i-\omega^2}))} }\right)^{1/2}.
\ee
 Now it is easy to see that the only isometry $U$ between ${\cal{K}_+}$ into ${\cal{K}_-}$ 
 can be defined through its action over $\phi_+$ as 
 $U \phi_+  \rightarrow e^{2\pi i \nu} \phi_-$, with $\nu$ a real parameter
 in the interval $[0,1)$. In particular it is a partial isometry and since the
 extensions of operators are in one to one correspondence with partial isometries
 the general theory of linear operators asserts that the only extensions that 
 $H_{rt}$ does admit are then
\be\label{Hext}
H_\nu = t \frac{d}{dt} t \frac{d}{dt}  + \lambda e^{\frac{\theta}{2} k} t 
     +  \omega^2,  
\ee
 with domain
\be\label{domHext}
D(H_\nu) := \left\{ \phi + \beta \phi_+  +  \beta e^{2\pi i \nu} \phi_- |
              \phi \in D(H_{rt}), \beta \in  {\mathbb{C}} \right\}.
\ee
 Here $D(H_{rt})$ is the domain of the symmetric operator $H_{rt}$ as defined
 in (\ref{symm}). Thus we have a one-parameter family of extensions of our
 original operator, labeled by the real number $\nu$. Now, since the domain
 of the partial isometry $U$ has been taken to be all of ${\cal{K}_+}$, the
 deficiency indices of these extensions $H_\nu$ are $n_\pm(H_\nu) = n_\pm(H_{rt})
 - {\textrm{dim}}({\cal{K}_+}) = 0$, which shows that we have achieved to construct
 a one-parameter family of self-adjoint extensions of $H_{rt}$.

 We are now ready to determine the eigenstates of our self-adjoint operator
 $H_\nu$. Let us denote by $\Omega$ the eigenvalues of $H_\nu$
\be\label{eigen}
H_\nu \phi(t) = \Omega \phi(t).
\ee
 The solutions to this differential equation are Bessel functions
 $J_{2\sqrt{\Omega-\omega^2}} \left[ 2\sqrt{\lambda} 
         e^{\frac{k\theta}{4}} t^{\frac{1}{2}} \right]$ and 
 $Y_{2\sqrt{\Omega-\omega^2}} \left[ 2\sqrt{\lambda} 
         e^{\frac{k\theta}{4}} t^{\frac{1}{2}} \right]$.  However
 only the Bessel functions of the first kind are square integrable
 on the interval $(0,\infty)$, provided $\Omega>\omega^2$, as can be checked 
 explicitly from the asymptotic expressions above. Of these, only certain
 values of $\Omega$ will render states that belong to the domain of $H_\nu$.
 To determine these values we use the fact that it is necessary for $\phi$ 
 to fulfill the condition
\be\label{cond}
\langle H^*_\nu \phi | \psi \rangle = \langle \phi | H_\nu \psi \rangle
\ee
 for all $\psi \in D(H_\nu)$. An equivalent form of this expression is given by
\be\label{econd}
\left[ t \psi \frac{d}{dt} \phi^* - 
        t \phi^* \frac{d}{dt} \psi \right]{\Bigg|_0^\infty} = 0.
\ee
 In particular for the linear combination $\psi=\phi_+ + e^{2\pi i \nu} \phi_-$ and using
 the asymptotic expressions for $J_k$ given as above we obtain the relation
\be\label{rel}
e^{2\pi i \nu} \sin\big[\pi(\sqrt{-i-\omega^2} - \sqrt{\Omega-\omega^2})\big]
 + \sin \big[\pi(\sqrt{i-\omega^2} - \sqrt{\Omega-\omega^2})\big] = 0.
\ee
 The real and imaginary parts of this expression are the same equation,
 solving for $\Omega$ we obtain
\be\label{solO}
\sqrt{\Omega - \omega^2} = \kappa + n,
\ee
 where $n$ is an integer and 
\be\label{solO2}
\kappa = {\textrm{Re}}(\sqrt{i-\omega^2}) 
   - \frac{1}{\pi} \tan^{-1} \left[ \frac{\cos(2\pi \nu)-1}{\sin(2\pi\nu)} 
      \tanh(\pi {\textrm{Im}}(\sqrt{i-\omega^2}) ) \right].  
\ee
 In summary, the spectrum of $H_\nu$ exists for
 $\Omega > \omega^2$, it is discrete and the eigenvalues $\Omega(n,\nu)$
 are parametrized by a real number $\nu \in [0,1)$. The normalized
 eigenstates are
\be\label{eigenstates}
u_n(x^0) = (2 (\kappa + n))^{1/2} J_{2(\kappa+n)} 
 \left[ 2\sqrt{\lambda} e^{\frac{1}{2}(x^0 + \frac{k\theta}{2})}
                 \right], 
\ee
 where we have returned to the original coordinate $x^0$.

 We see that the parameter $\theta$ induces a shift
 in the minisuperspace variable $x^0$ proportional to $k$. 
 The full minisuperspace eigenfunctions  are then 
 $\psi_n(x) = \exp \big( i {\bf p} \cdot {\bf x} \big) u_n(x^0)$. 
 In order to visualize more clearly the effect of noncommutativity, we construct 
 wave-packets by summing over momentum states assuming that  
 $k$ is, say $p_1$ and the rest of the $p_j$ vanish. 
 Denoting by $\Psi(x^0,x^1)$  the full 
 eigenfunctions under these assumptions,
 the formal structure of the wave-packet under consideration is
\be\label{wavepacket}
\Phi(x^0, x^1) =  \int_{-\infty}^\infty dk e^{-\alpha(k-\beta)^2} 
     \Psi(x^0,x^1),  
\ee 
 where we have introduced
 a gaussian weighting function with parameters $\alpha$ and $\beta$.
 The integral is performed numerically keeping the order $2(\kappa+n)$
 of the Bessel functions a positive real number, since $\kappa$ depends
 on $k$ through $\omega^2$ and we have to fullfil the condition $\Omega>\omega^2$.
 For simplicity we take $\nu=0$, in such case $\kappa$ remains in the interval
 $(0,1 / \sqrt{2})$ as $k$ varies along the whole real line,
 as is easily seen from the expression (\ref{solO2})
 {\footnote{Here we are taking into account only the positive square root, and
 also we neglect the term $N-1$ so that actually $\omega^2=k^2$.}}.
 In figure 2
 we show a comparison
 between the ordinary ($\theta = 0$) and the noncommutative case of the
 probability density $|\Phi|^2$ in the $(x^0,x^1)$ subspace.
 We observe an overall shift in the $x^0$ direction
 of the absolute maximum of the wave-packet, which is accompanied by a large
 enhancement of the height of the peak of the packet.

\begin{figure}\label{fig:rollfig1}
\let\picnaturalsize=N
\def\picsize{18.0 cm}
\def\picfilename{fig1.eps}
\ifx\nopictures Y\else{\ifx\epsfloaded Y\else\input epsf \fi
\let\epsfloaded=Y
\centerline{\ifx\picnaturalsize N\epsfxsize \picsize\fi \epsfbox{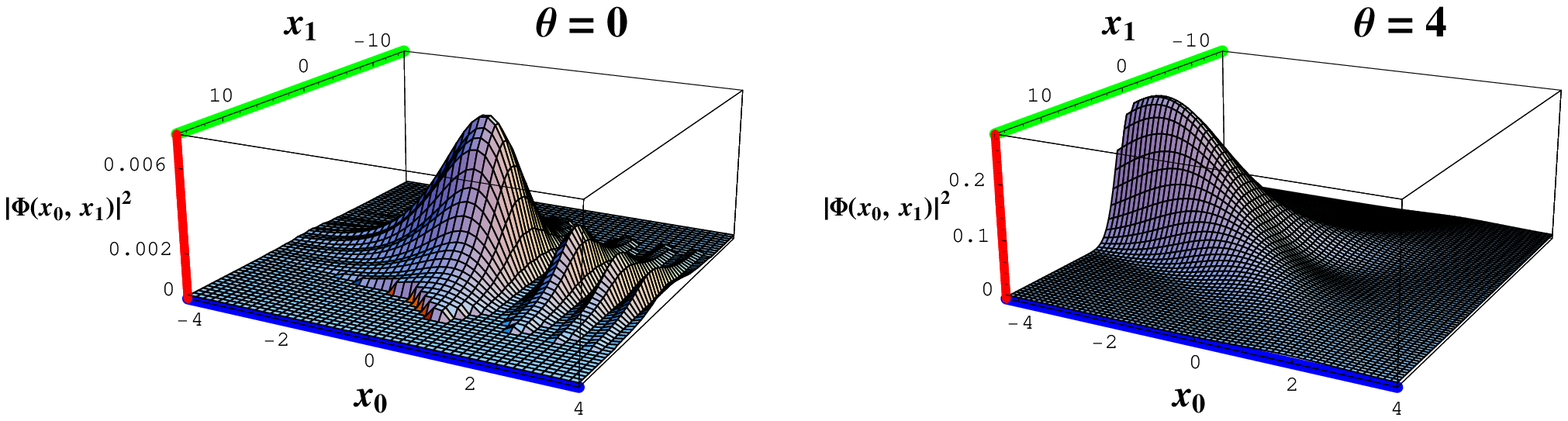}}}\fi
{\small {Figure 2.  Wave packet for the half S-brane. Here we take $n=0$ and the order of the
           eigenstates  $2\kappa$ is kept a positive real number. 
        The values of the remaining parameters are $\nu=0$ and $\alpha=\beta=1$. }}
\end{figure}

\section{The S-brane}\label{sbranesecc}

As mention in the introduction, the worldsheet action for the bouncing tachyon
is given by equation (\ref{worldsheet}). The corresponding minisuperspace 
noncommutative hamiltonian $H_{bt}$ is easily shown to be
\be\label{btham}
H_{bt} = \partial_{x^0}^2 + 2\lambda \cosh{\left( x^0 + \frac{1}{2} \theta k \right)} 
         + \omega^2.
\ee
Since the formal arguments in the previous section apply straightforwardly
to this hamiltonian, we shall be rather brief. Thus $H_{bt}$ becomes a 
symmetric operator if we choose for its domain of definition the following one
\be\label{symmbt}
D(H_{bt}) := \left\{ \psi | \psi \in L^2(-\infty,\infty), H \psi \in L^2(-\infty,\infty),
 \lim_{t \rightarrow -\infty, \infty}   \frac{d\psi(x^0)}{dx^0} = 0 
\right\}.
\ee
The adjoint operator $H^*_{bt}$ takes the same differential form as $H_{bt}$ 
and we now seek the kernel subspaces ${\cal{K}_\pm}$ of the corresponding
operators $(H^*_{bt} \pm i)$. The solutions of the equations $H^*_{bt} \psi = \pm i \psi$
are known as Mathieu functions $se_\nu(z,\lambda)$ and $ce_\nu(z,\lambda)$ 
of imaginary argument (sometimes called 
modified Mathieu functions). For each sign there are two linearly independent
solutions, both of which are square integrable as we shall see later on. 
Therefore the relevant subspaces are
\be\label{kermat+}
{\cal {K}_+} = \left\{ c_1 \phi^+_1(x^0) 
                     + c_2 \phi^+_2(x^0) 
| c_1, c_2 \in {\mathbb{C}}
 \right\},
\ee
\be\label{kermat-}
{\cal {K}_-} = \left\{ c_1 \phi^-_1(x^0)
                     + c_2 \phi^-_2(x^0)
| c_1, c_2 \in {\mathbb{C}},
 \right\}
\ee
where
\be\label{matfi+1}
\phi^+_1(x^0) = se_{4(i-\omega^2)}
\left( -\frac{1}{2} i \left( x^0 + \frac{1}{2} \theta k \right), 4\lambda \right),
\ee
\be\label{matfi+2}
\phi^+_2(x^0) = ce_{4(i-\omega^2)}
\left( -\frac{1}{2} i \left( x^0 + \frac{1}{2} \theta k \right), 4\lambda \right)
\ee
and
\be\label{matfi-1}
\phi^-_1(x^0) = se_{4(-i-\omega^2)}
\left( -\frac{1}{2} i \left( x^0 + \frac{1}{2} \theta k \right), 4\lambda \right),
\ee
\be\label{matfi-2}
\phi^-_2(x^0) = ce_{4(-i-\omega^2)}
\left( -\frac{1}{2} i \left( x^0 + \frac{1}{2} \theta k \right), 4\lambda \right),
\ee
with the resultant deficiency indices of $H_{bt}$ being $(n_+,n_-) = (2,2)$. From 
this we see that in order to construct self-adjoint extensions of the operator $H_{bt}$ 
we need to find all the possible partial isometries from ${\cal {K}_+}$ into 
${\cal {K}_-}$ such that their domain of definition has dimension $2$. We therefore
choose the domain of definition $I(U) \subseteq {\cal {K}_+}$ of a possible 
isometry $U$ to be ${\cal {K}_+}$ itself, then it is easy to see that we only 
have two possible families of partial isometries, which can be defined as
\be\label{piso1}
U_1 \left( { \frac{1}{\| \phi^+_1 \|} \phi^+_1   \atop  
             \frac{1}{\| \phi^+_2 \|} \phi^+_2        }  \right) 
   \longrightarrow 
  e^{2\pi i \nu_1} 
    \left( { \frac{1}{\| \phi^-_1 \|} \phi^-_1   \atop  
             \frac{1}{\| \phi^-_2 \|} \phi^-_2        }  \right) 
\ee
and
\be\label{piso2}
U_2 \left( { \frac{1}{\| \phi^+_1 \|} \phi^+_1   \atop  
             \frac{1}{\| \phi^+_2 \|} \phi^+_2        }  \right) 
   \longrightarrow 
  e^{2\pi i \nu_2} 
    \left( { \frac{1}{\| \phi^-_2 \|} \phi^-_2   \atop  
             \frac{1}{\| \phi^-_1 \|} \phi^-_1        }  \right), 
\ee
with $\nu_1$ and $\nu_2$ real parameters in the interval $[0,1)$. 
Moreover, it turns out that the two resultant self-adjoint extensions 
of $H_{bt}$ associated with $U_1$ and $U_2$ are really equivalent, since we can 
readily check that their domains of definition $D(H_{U_1})$ and $D(H_{U_2})$ 
are the same set. In this way, the only self-adjoint extension of $H_{bt}$ is given by
\be\label{bthamext}
H_\nu = \partial_{x^0}^2 + 2\lambda \cosh{\left( x^0 + \frac{1}{2} \theta k \right)} 
         + \omega^2,
\ee
with domain
\bea\label{dombt}
D(H_\nu) &:=& \left\{   \phi +  c_1    \frac{1}{\| \phi^+_1 \|} \phi^+_1 
                           +  c_2    \frac{1}{\| \phi^+_2 \|} \phi^+_2 
         + e^{2\pi i \nu} \left( 
                              c_1    \frac{1}{\| \phi^-_1 \|} \phi^-_1   
                           +  c_2    \frac{1}{\| \phi^-_2 \|} \phi^-_2   
 \right)  \right. \nonumber \\ 
& &  \qquad\qquad | \phi \in D(H_{bt}), c_1, c_2 \in {\mathbb{C}}     
 {\Big{\}}}.
\eea

Now the two linearly independent solutions of the eigenvalue equation 
$H_\nu \phi(x^0) = \Omega \phi(x^0)$ are 
$se_{4(\Omega-\omega^2)}
\left( -\frac{1}{2} i \left( x^0 + \frac{1}{2} \theta k \right), 4\lambda \right)$ 
and 
$ce_{4(\Omega-\omega^2)}
\left( -\frac{1}{2} i \left( x^0 + \frac{1}{2} \theta k \right), 4\lambda \right)$, 
of these, only those which satisfy the condition
\be\label{conbt}
\left[  \psi \frac{d}{dx^0} \phi^* 
  -  \phi^* \frac{d}{dx^0} \psi  \right] {\Bigg |}^\infty_{-\infty} = 0,
\ee
for all $\psi \in D(H_\nu)$, will be elements of the domain of $H_\nu$ and
therefore eigenfunctions of it. At this point it is useful to obtain expressions
for the asymptotic behavior of the Mathieu functions as $x^0 \rightarrow \pm \infty$.
A quick argument which we shall use was suggested in \cite{Fredenhagen:2003ut}, we
start by noticing that the operator (\ref{btham}) has the following asymptotics
\be\label{hbtasympt}
H_{bt} {\longrightarrow_{x^0 \rightarrow \pm\infty}} \qquad
\partial_{\xi}^2 + \lambda e^{\pm\frac{\theta}{2} k} e^{\xi} 
     +  \omega^2
\ee
where
\be
\xi := \left\{ {  {x^0, \ \ \ \  \mbox{for } x^0 > 0} 
          \atop {-x^0, \ \ \ \ \ \ \mbox{for } x^0 < 0} } \right.
\ee
From this we see that the solutions of the differential operator $H_{bt}$
must have the same aymptotics as the solutions of the differential operators
in the rhs of equation (\ref{hbtasympt}) as $\xi \rightarrow \infty$, in 
particular
\be
se_{q^2}
\left( -\frac{1}{2} i \left( x^0 + \frac{1}{2} \theta k \right), \lambda \right)
{\longrightarrow_{x^0 \rightarrow \pm\infty}} \qquad
Y_q \left[  \sqrt{\lambda} e^{\frac{1}{2}\left( \xi \pm \frac{1}{2} \theta 
k \right)}  \right]
\ee
and
\be
ce_{q^2}
\left( -\frac{1}{2} i \left( x^0 + \frac{1}{2} \theta k \right), \lambda \right)
{\longrightarrow_{x^0 \rightarrow \pm\infty}} \qquad
J_q \left[  \sqrt{\lambda} e^{\frac{1}{2}\left( \xi \pm \frac{1}{2} \theta 
k \right)}  \right],
\ee
where $\xi$ is defined as above. Now, for our purposes, it suffices to take
\be
\psi = 
     \phi^+_1  +   \phi^+_2 
         + e^{2\pi i \nu} \left(  \phi^-_1  +  \phi^-_2   
 \right)
\ee
and
\be
\phi = se_{4(\Omega-\omega^2)}
\left( -\frac{1}{2} i \left( x^0 + \frac{1}{2} \theta k \right), 4\lambda \right)
+ ce_{4(\Omega-\omega^2)}
\left( -\frac{1}{2} i \left( x^0 + \frac{1}{2} \theta k \right), 4\lambda \right)
\ee
and, upon substitution in (\ref{conbt}) and making use of the above asymptotics
together with the corresponding ones for the Bessel functions, we are able to show
that condition (\ref{conbt}) is satisfied automatically for any value of $\Omega$,
which of course means that the spectrum of the s-brane is continuous.

Summing up, the full minisuperspace eigenfunctions of the s-brane are given by
\be
\psi_\Omega(x) = \exp \big( i {\bf p} \cdot {\bf x}  \big)
se_{4(\Omega-\omega^2)}
\left( -\frac{1}{2} i \left( x^0 + \frac{1}{2} \theta k \right), 4\lambda \right)
\ee
and 
\be
\psi_\Omega(x) = \exp \big( i {\bf p} \cdot {\bf x} \big)
ce_{4(\Omega-\omega^2)}
\left( -\frac{1}{2} i \left( x^0 + \frac{1}{2} \theta k \right), 4\lambda \right), 
\ee
with $\Omega$ being the real eigenvalue.

\section{Summary and Conclusions}\label{sum}

In this paper we have studied some consequences of the noncommutative deformation of the
minisuperspace approach of the CFT construction of S-branes describing 
the bouncing and rolling tachyon. This
discussion is done in the spirit of Ref. \cite{Garcia-Compean:2001wy} where for the case
of quantum cosmology an exact solution to the Moyal deformation of Wheeler-DeWitt
equation for the Kantowski-Sachs metric was found.

Similarly to the quantum cosmology case, for the CFT approach of the bouncing and rolling tachyon
we find exact solutions of the corresponding hamiltonian constraint given by the Klein-Gordon
equation in the minisuperspace, written in terms of elementary functions with the argument
modified by the noncommutativity parameter $\theta$. We found how this resulting effective
equation for the corresponding energy spectrums have analytic exact solution without the need for
an expansion in the noncommutativity parameter $\theta$. The resulting eigenstates of the
corresponding models acquire a shift in the timelike parameter proportional to $\theta$. For the
purpose of visualizing the effect of noncommutativity, wave packets with suitable weighing
functions are constructed and analyzed for especial cases. In general the $\theta$-effect shows
up as an enhancement of the localization in the timelike direction of the wave packet, apart from
an overall shift.

We would like to mention that comparison with the results of the commutative case
treated in reference \cite{Fredenhagen:2003ut} is not directly applicable, since the
authors' approach to the subject involves, in particular for the s-brane, imposing independent 
boundary conditions at the far past and at the far future, which traduces itself in having
two independent parameters ($\nu_\pm$ instead of our $\nu$ in equation (\ref{dombt})).
This in no way implies that our results in the $\theta \rightarrow 0$ limit are contradictory, 
rather they are complementary 
since it is well known that general symmetric operators can have several independent
self-adjoint extensions, differing only in subtle changes in the physical system being
described, and this usually has to do with the chosen boundary conditions. Our approach
to the subject has been performed through the slightly more formal procedures of reference
\cite{Kobayashi:1996kg}, and as it was to be expected, in the commutative limit our results
agree with those there, where applicable.

In a different order of ideas, as is evident from the discussion of the half s-brane in the main
body of the text, the effect of the introduction of the noncommutative parameter $\theta$ can be
naively accounted for with the substitution $ \lambda \rightarrow e^{\frac{1}{2} k \theta}
\lambda $, in the commutative action, where $k$ is defined as in equation (\ref{rolling}). Thus,
in principle, we could try to analyze the known results of the commutative case, with the naive
substitution above. The possibility of having a slight wider window of options due to the
enlarged parameter space with the inclusion of $\theta$, is an interesting motivation for
pursuing this idea. Finally, it would be interesting to give a derivation of the noncommutative
minisuperspace from a CFT description of bouncing and rolling tachyon by including the background
$B$-field term, in a similar spirit of how noncommutativity is usually derived from string
theory. Progress along these lines will be posted elsewhere.

\vskip 2truecm
\centerline{\bf Acknowledgments}
This work was supported in part by CONACyT M\'exico Grant 33951E. H. G.-C. thanks ICTP for
hopitality as an ICTP-Associate Member. J. G.-B. is supported by
a CONACyT graduate fellowship.


\end{document}